*Research Article*

# Linear Sigma Model at Finite Temperature and Baryonic Chemical Potential Using the $N$-Midpoint Technique

### M. Abu-Shady

*Department of Mathematics, Faculty of Science, Menoufia University, Shiben El-Koom, 32511, Egypt*

Correspondence should be addressed to M. Abu-Shady; dr.abushady@gmail.com

A baryonic chemical potential ($\mu_b$) is included in the linear sigma model at finite temperature. The effective mesonic potential is numerically calculated using the $N$-midpoint rule. The meson masses are investigated as functions of the temperature ($T$) at fixed value of baryonic chemical potential. The pressure and energy density are investigated as functions of temperature at fixed value of $\mu_b$. The obtained results are in good agreement in comparison with other techniques. We conclude that the calculated effective potential successfully predicts the meson properties and thermodynamic properties at finite baryonic chemical potential.

## 1. Introduction

Quantum Chromodynamics (QCD) is a fundamental theory of strong interactions since it is renormalizable. However, thermodynamics of QCD is not well understood because of its nonperturbative nature. In particular, QCD phase diagram is an essential for understanding not only natural phenomena such as compact stars and the early universe but also laboratory experiments such as relativistic heavy-ion collisions. The quantitative calculations of phase diagram from first-principle lattice QCD (LQCD) have the well-known sign problem when the baryon chemical potential is included ($\mu_B$) [1].

At nonzero baryonic chemical potential, standard numerical lattice simulations do not work well. Therefore our knowledge of the QCD phase diagram at nonzero $T$ and $\mu_b$ relies exclusively on effective models. One of the effective models for describing baryon properties is the linear sigma model, which was suggested by Gell-Mann and Levy [2] to describe nucleons interacting via sigma ($\sigma$) and pion ($\pi$) exchanges. The model is extended to describe the quark interactions by Birse and Banerjee [3]. The model is modified to provide good description of baryon properties at zero temperature as in [4–9]. At finite temperature and zero baryonic potential, the model gives a good description of meson properties and thermodynamic properties as in [10–17].

The baryonic chemical potential is studied in the linear sigma model at finite temperature as in [18–20]. Also, the isospin chemical potential is studied in the same model by using the Cornwall-Jackiw-Tomboulis formalism [21].

The aim of this work is to calculate the effective mesonic potential in the presence of the baryonic chemical potential at finite temperature using the $N$-midpoint rule. The present paper is extended from the previous work [17] at finite baryon chemical potential. The calculated effective potential is used to obtain meson properties, the pressure, and the energy density.

This paper is organized as follows. In Section 2, the linear sigma model at zero temperature and baryonic chemical potential is briefly explained. In Section 3, the effective mesonic potential is presented at finite chemical potential. Numerical calculations and results are discussed in Section 4. Summary and conclusion are presented in Section 5.

## 2. The Linear Sigma Model

The interactions of quarks via the exchange of $\sigma$- and $\pi$-meson fields is given by the Lagrangian density [3] as follows:

$$L(r) = i\overline{\Psi}\partial_\mu\gamma^\mu\Psi + \frac{1}{2}\left(\partial_\mu\sigma\partial^\mu\sigma + \partial_\mu\boldsymbol{\pi}\cdot\partial^\mu\boldsymbol{\pi}\right) \\ + g\overline{\Psi}\left(\sigma + i\gamma_5\boldsymbol{\tau}\cdot\boldsymbol{\pi}\right)\Psi - U^{T(0)}(\sigma,\pi), \tag{1}$$

with

$$U^{T(0)}(\sigma, \pi) = \frac{\lambda^2}{4}\left(\sigma^2 + \pi^2 - \nu^2\right)^2 + m_\pi^2 f_\pi \sigma. \tag{2}$$

$U^{T(0)}(\sigma, \pi)$ is the meson-meson interaction potential where $\Psi$, $\sigma$ and $\pi$ are the quark, sigma, and pion fields, respectively. In the mean-field approximation, the meson fields are treated as time-independent classical fields. This means that we replace the power and products of the meson fields by corresponding powers and the products of their expectation values. The meson-meson interactions in (2) lead to hidden chiral $SU(2) \times SU(2)$ symmetry with $\sigma(r)$ taking on a vacuum expectation value

$$\langle \sigma \rangle = -f_\pi, \tag{3}$$

where $f_\pi = 93$ MeV is the pion decay constant. The final term in (2) is included to break the chiral symmetry explicitly. It leads to the partial conservation of axial-vector isospin current (PCAC). The parameters $\lambda^2$, $\nu^2$ can be expressed in terms of $f_\pi$, meson masses, as

$$\begin{aligned} \lambda^2 &= \frac{m_\sigma^2 - m_\pi^2}{2 f_\pi^2}, \\ \nu^2 &= f_\pi^2 - \frac{m_\pi^2}{\lambda^2}. \end{aligned} \tag{4}$$

## 3. The Linear Sigma Model at Finite Temperature and Baryonic-Chemical Potential

The purpose of this section is to include the finite temperature and chemical potential in (1); the effective potential is extended to include the chiral interacting meson fields with quarks fields at the finite temperature and baryonic chemical potential [18]. As follows:

$$\begin{aligned} &U_{\text{eff}}(\sigma, \pi, T, \mu_b) \\ &= U^{T(0)}(\sigma, \pi) \\ &\quad - 12 \int \frac{d^3 p}{(2\pi)^3} \\ &\qquad \times \Bigg[ T \ln\left(\exp\left(\mu_b - \frac{1}{T}\sqrt{p^2 + g^2(\sigma^2 + \pi^2)}\right) + 1\right) \\ &\qquad\quad + T \ln\left(\exp\left(-\mu_b - \frac{1}{T}\sqrt{p^2 + g^2(\sigma^2 + \pi^2)}\right) + 1\right) \Bigg]. \end{aligned} \tag{5}$$

The first term is the potential in the tree level defined in (2) and the second term is the chiral meson fields interacting with quarks at the finite temperature and baryonic chemical potential. Nonzero values of the chiral fields in the chiral broken phase dynamically generate a quark mass $m_q^2 = g^2\langle \sigma^2 + \pi^2 \rangle$. The integration is taken over momentum space ($p$) (for details, see [18]).

*3.1. Numerical Calculations by Using the N-Midpoint Technique.* The purpose of this subsection is to calculate the effective mesonic potential, the $\sigma$ and $\pi$-meson masses, the pressure, and the energy density using the $N$-midpoint technique. We rewrite (5) as follows:

$$\begin{aligned} &U_{\text{eff}}(\sigma, \pi, T, \mu_b) \\ &= U^{T(0)}(\sigma, \pi) \\ &\quad - \frac{6T}{\pi^2}\int_0^\infty p^2 \\ &\qquad \times \ln\left(\exp\left(\mu_b - \frac{1}{T}\sqrt{p^2 + g^2(\sigma^2 + \pi^2)}\right) + 1\right)dp \\ &\quad - \frac{6T}{\pi^2}\int_0^\infty p^2 \\ &\qquad \times \ln\left(\exp\left(-\mu_b - \frac{1}{T}\sqrt{p^2 + g^2(\sigma^2 + \pi^2)}\right) + 1\right)dp. \end{aligned} \tag{6}$$

Equation (6) is written in the dimensionless form as follows:

$$\begin{aligned} &U_{\text{eff}}(\sigma, \pi, T, \mu_b) \\ &= f_\pi^4 \Bigg[ U^{T(0)}(\sigma', \pi') - \frac{6T'}{\pi^2} \\ &\qquad \times \int_0^\infty p'^2 \\ &\qquad \times \Big[\ln\left(\exp\left(\mu_b' - \frac{1}{T'}\sqrt{P'^2 + g^2(\sigma'^2 + \pi'^2)}\right) + 1\right) \\ &\qquad\quad - \ln\Big(\exp\left(-\mu_b' - \frac{1}{T'}\sqrt{P'^2 + g^2(\sigma'^2 + \pi'^2)}\right) \\ &\qquad\qquad\qquad + 1\Big)\Big] dp' \Bigg], \end{aligned} \tag{7}$$

with

$$U^{T(0)}(\sigma', \pi') = \frac{\lambda^2}{4}\left(\sigma'^2 + \pi'^2 - \nu'^2\right)^2 + m_\pi^2 \sigma', \tag{8}$$

where

$$\begin{aligned} p' &= \frac{p}{f_\pi}, & \sigma' &= \frac{\sigma}{f_\pi}, & \pi' &= \frac{\pi}{f_\pi}, \\ T' &= \frac{T}{f_\pi}, & \nu' &= \frac{\nu}{f_\pi}, & m_\pi' &= \frac{m_\pi}{f_\pi}, & \mu' &= \frac{\mu}{f_\pi}. \end{aligned} \tag{9}$$

The $U^{T(0)}(\sigma', \pi')$ is the dimensionless form of $U^{T(0)}(\sigma, \pi)$. Substituting $p' = -\ln y$ into (7), we obtain

$$\begin{aligned} &U_{\text{eff}}(\sigma, \pi, T, \mu_b) \\ &= f_\pi^4 \Bigg[ U^{T(0)}(\sigma', \pi') - \frac{6T'}{\pi^2} \\ &\qquad \times \int_0^1 \frac{(\ln y)^2}{y} \\ &\qquad \times \Bigg[ \ln\Bigg(\exp\bigg(\mu_b \\ &\qquad\qquad\qquad - \frac{1}{T'}\sqrt{(\ln y)^2 + g^2(\sigma'^2 + \pi'^2)}\bigg) + 1\Bigg) \end{aligned}$$

$$+ \ln \left( \exp \left( - \mu_b \right. \right.$$
$$\left. - \frac{1}{T'} \sqrt{(\ln y)^2 + g^2 \left( \sigma'^2 + \boldsymbol{\pi}'^2 \right)} \right)$$
$$\left. \left. + 1 \right) \right] dy \Bigg],$$

(10)

and we write dimensionless form of $U_{\text{eff}}(\sigma', \boldsymbol{\pi}', T', \mu_b')$ as follows:

$$U_{\text{eff}} \left( \sigma', \boldsymbol{\pi}', T', \mu_b' \right)$$
$$= U^{T(0)} \left( \sigma', \boldsymbol{\pi}' \right) - \frac{6T'}{\pi^2}$$
$$\times \int_0^1 \frac{(\ln y)^2}{y}$$
$$\times \left[ \ln \left( \exp \left( \mu_b' - \frac{1}{T'} \sqrt{(\ln y)^2 + g^2 \left( \sigma'^2 + \boldsymbol{\pi}'^2 \right)} \right) + 1 \right) \right.$$
$$\left. + \ln \left( \exp \left( -\mu_b' - \frac{1}{T'} \sqrt{(\ln y)^2 + g^2 \left( \sigma'^2 + \boldsymbol{\pi}'^2 \right)} \right) + 1 \right) \right] dy.$$

(11)

Using midpoint rule (for details, see [22, 23]), we obtain the approximated integral as follows:

$$U_{\text{eff}} \left( \sigma, \boldsymbol{\pi}, T, \mu_b \right)$$
$$= f_\pi^4 \left[ U^{T(0)} \left( \sigma', \boldsymbol{\pi}' \right) - \frac{6T'}{\pi^2} A \right.$$
$$\times \ln \left( \exp \left( \mu_b' - \frac{1}{T'} \sqrt{g^2 \left( \sigma'^2 + \boldsymbol{\pi}'^2 \right) + B} \right) + 1 \right)$$
$$+ \frac{6T'}{\pi^2} A \ln \left( \exp \left( - \mu_b' \right. \right.$$
$$\left. \left. \left. - \frac{1}{T'} \sqrt{g^2 \left( \sigma'^2 + \boldsymbol{\pi}'^2 \right) + B} \right) + 1 \right) \right],$$

(12)

where

$$A = \frac{1}{n} \sum_{i=0}^n \frac{1}{(1/n) i + 1/2n} \ln^2 \left( \frac{1}{n} i + \frac{1}{2n} \right),$$
$$B = \sum_{i=0}^n \ln^2 \left( \frac{1}{n} i + \frac{1}{2n} \right).$$

(13)

In [12, 17], the authors applied the second derivation of the effective potential respect to $\sigma'$ and $\pi'$ to obtain the effective meson masses. The first derivative of the effective potential $U_{\text{eff}}(\sigma', \pi', T')$ is given by

$$\frac{\partial U_{\text{eff}} \left( \sigma', \boldsymbol{\pi}', T, \mu_b' \right)}{\partial \sigma}$$
$$= \frac{1}{f_\pi} \frac{\partial U_{\text{eff}} \left( \sigma', \boldsymbol{\pi}', T, \mu_b' \right)}{\partial \sigma'}$$

$$= f_\pi^3 \left[ \frac{\partial U_0 \left( \sigma', \boldsymbol{\pi}' \right)}{\partial \sigma'} - \frac{6T'}{\pi^2} \frac{d_1 d_2}{(d_3 + d_3 d_2)} \right.$$
$$\left. - \frac{6T'}{\pi^2} \frac{d_1 d_4}{(d_3 + d_3 d_4)} \right],$$

(14)

where

$$d_1 = -A g^2 \sigma',$$
$$d_2 = \exp \left( \mu_b' - \frac{1}{T'} \sqrt{B + g^2 \left( \sigma'^2 + \boldsymbol{\pi}'^2 \right)} \right),$$
$$d_3 = T' \sqrt{B + g^2 \left( \sigma'^2 + \boldsymbol{\pi}'^2 \right)},$$
$$d_4 = \exp \left( -\mu_b' - \frac{1}{T'} \sqrt{B + g^2 \left( \sigma'^2 + \boldsymbol{\pi}'^2 \right)} \right).$$

(15)

Then, we obtain the effective sigma mass as follows:

$$M_\sigma^2 (T) = \frac{\partial^2 U_{\text{eff}} \left( \sigma', \boldsymbol{\pi}', T, \mu_b' \right)}{f_\pi^2 \partial \sigma'^2},$$
$$M_\sigma (T) = f_\pi \left[ \frac{\partial^2 U_0 \left( \sigma', \boldsymbol{\pi}' \right)}{\partial \sigma'^2} - \frac{6T'}{\pi^2} \frac{\partial}{\partial \sigma'} \right.$$
$$\left. \times \left( \frac{d_1 d_2}{d_3 + d_3 d_2} + \frac{d_1 d_4}{d_3 + d_3 d_4} \right) \right]^{1/2},$$

(16)

where

$$\frac{\partial}{\partial \sigma'} \left( \frac{d_1 d_2}{d_3 + d_3 d_2} \right) = d_1 \frac{d_2'}{d_3 + d_2 d_3} + d_2 \frac{d_1'}{d_3 + d_2 d_3}$$
$$- \frac{d_1 d_2}{(d_3 + d_2 d_3)^2} \left( d_3' + d_2 d_3' + d_3 d_2' \right),$$
$$\frac{\partial}{\partial \sigma'} \left( \frac{d_1 d_4}{d_3 + d_3 d_4} \right) = d_1 \frac{d_4'}{d_3 + d_4 d_3} + d_4 \frac{d_1'}{d_3 + d_4 d_3}$$
$$- d_1 \frac{d_4}{(d_3 + d_4 d_3)^2} \left( d_3' + d_4 d_3' + d_3 d_4' \right).$$

(17)

We obtain the derivation of $d_1$, $d_2$, and $d_3$ with respect to $\sigma'$ as follows:

$$d_1' = -A g^2,$$
$$d_2' = \frac{-g^2 \sigma'}{T' \sqrt{B + g^2 \left( \sigma'^2 + \boldsymbol{\pi}'^2 \right)}}$$
$$\times \exp \left( \mu_b' - \frac{1}{T'} \sqrt{B + g^2 \left( \sigma'^2 + \boldsymbol{\pi}'^2 \right)} \right),$$
$$d_4' = \frac{-g^2 \sigma'}{T' \sqrt{B + g^2 \left( \sigma'^2 + \boldsymbol{\pi}'^2 \right)}}$$

$$\times \exp\left(-\mu_b' - \frac{1}{T'}\sqrt{B + g^2\left(\sigma'^2 + \pi'^2\right)}\right),$$

$$d_3' = \frac{\sigma' g^2 T}{\sqrt{B + g^2\left(\sigma'^2 + \pi'^2\right)}}.$$

(18)

Similarly, we obtain the effective pion mass as follows:

$$M_\pi(T) = f_\pi \left[\frac{\partial^2 U(\sigma', \pi')}{\partial \pi'^2} - \frac{6}{\pi^2} T' \frac{\partial}{\partial \pi'} \right.$$

$$\left. \times \left(\frac{d_5 d_2}{d_3 + d_3 d_2} + \frac{d_5 d_4}{d_3 + d_3 d_4}\right)\right]^{1/2},$$

(19)

where

$$d_5 = -Ag^2 \pi',$$

$$\frac{\partial}{\partial \pi'}\left(\frac{d_5 d_2}{d_3 + d_3 d_2}\right) = d_2 \frac{d_5'}{d_3 + d_2 d_3} + d_5 \frac{d_2'}{d_3 + d_2 d_3}$$

$$- d_2 \frac{d_5}{(d_3 + d_2 d_3)^2}\left(d_3' + d_2 d_3' + d_3 d_2'\right),$$

$$\frac{\partial}{\partial \pi'}\left(\frac{d_5 d_4}{d_3 + d_3 d_4}\right) = d_4 \frac{d_5'}{d_3 + d_4 d_3} + d_5 \frac{d_4'}{d_3 + d_4 d_3}$$

$$- d_4 \frac{d_5}{(d_3 + d_4 d_3)^2}\left(d_3' + d_4 d_3' + d_3 d_4'\right).$$

(20)

We obtain the derivation of $d_2$, $d_4$, and $d_5$ with respect to $\pi'$ as follows:

$$d_5' = -Ag^2,$$

$$d_2' = \frac{-g^2 \pi'}{T'\sqrt{B + g^2\left(\sigma'^2 + \pi'^2\right)}}$$

$$\times \exp\left(\mu_b' - \frac{1}{T'}\sqrt{B + g^2\left(\sigma'^2 + \pi'^2\right)}\right),$$

$$d_4' = \frac{-g^2 \pi'}{T'\sqrt{B + g^2\left(\sigma'^2 + \pi'^2\right)}}$$

$$\times \exp\left(-\mu_b' - \frac{1}{T'}\sqrt{B + g^2\left(\sigma'^2 + \pi'^2\right)}\right),$$

$$d_3' = \frac{\pi' g^2}{T'\sqrt{B + g^2\left(\sigma'^2 + \pi'^2\right)}}.$$

(21)

In [24], the local pressure is given in the nondimensional form

$$P(\sigma', \pi', T) = U^{T(0)}(\sigma', \pi') - U_{\text{eff}}(\sigma', \pi', T').$$

(22)

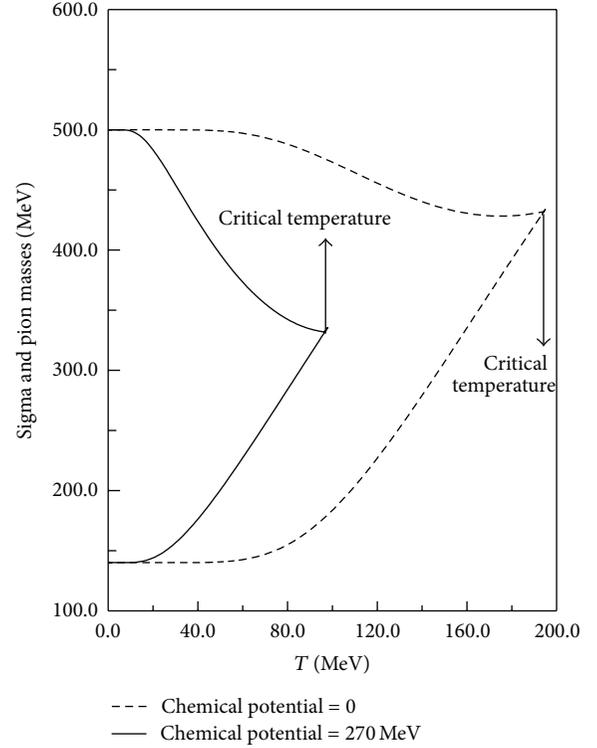

FIGURE 1: Sigma and pion masses are plotted as functions of temperature in the absence and presence of baryonic chemical potential.

Therefore, the energy density is defined as in [25] as follows:

$$E(T', \mu_b') = -P(T', \mu_b') + T' \frac{\partial P(T', \mu_b')}{\partial T'} + \mu_b' \frac{\partial P(T', \mu_b')}{\partial \mu_b'}.$$

(23)

## 4. Discussion of Results

In (6), the integration is solved by the $N$-midpoint algorithm as in [17]. The index $n$ is taken $n = 1000$ to get a good accuracy for the numerical integration. In the present work, the parameters of the model such as $m_\pi = 140$ MeV, $m_\sigma = 500 \rightarrow 700$ MeV, $f_\pi = 93$ MeV, and the coupling constant $g$ at zero temperature are used as the initial parameters at the finite temperature. The calculated effective potential in (12) is used to study the effect of finite temperature on the effective pion and sigma masses, the pressure, and the energy density in the presence of the baryonic chemical potential. In Figure 1, the sigma and pion masses are plotted as functions of temperature in the absence of baryonic chemical potential ($\mu_b = 0$) and the presence of the baryonic chemical potential $\mu_b = 270$ MeV. We note that the sigma mass decreases with increasing temperature while pion mass increases with increasing temperature. The curves crossed at the critical temperature are equal to 191 MeV which is in agreement with the previous work [17].

At finite baryonic chemical potential, it is more difficult to predict the phase transition using lattice QCD. Phenomenology models such as the quark sigma model and the NJL model are used to describe the phase transition. Chen et al. [26] pointed out that the prediction of the chiral phase transition at finite baryonic chemical potential is still under considerable speculation. In present work, the baryonic chemical potential taken is equal to $\mu_b = 270$ MeV. The qualitative features of sigma and pion masses are not changed by involving the baryonic potential while the critical temperature shifted from 191 MeV to 99 MeV by including the baryonic potential in the model. In comparison with Scavenius et al. [18], they found that the phase transition is first-order at coupling constant $g = 4.5$ and the phase transition changed to crossover by decreasing coupling constant. Therefore, the phase transition is crossover when they took the baryonic chemical potential $\mu_b = 270$ MeV in the Nambu-Jona-Lasinio (NJL) model and PNJL model. Hence, our results are in agreement with the results of [27].

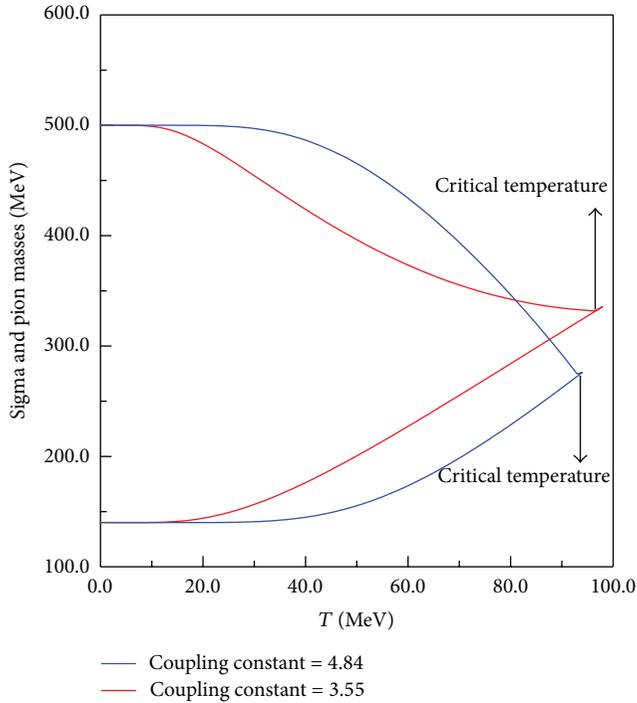

FIGURE 2: Sigma and pion masses are plotted as functions of temperature for two values of coupling constant $g$.

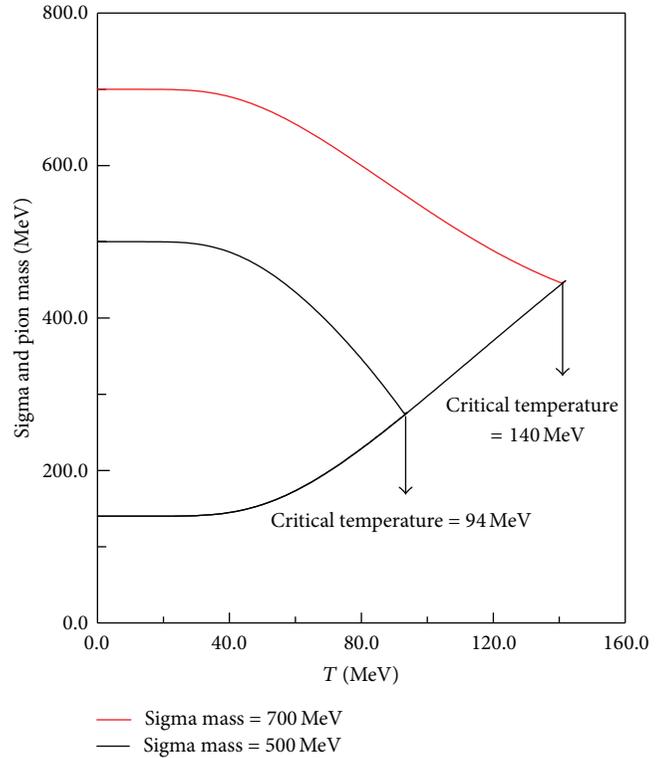

FIGURE 3: Sigma and pion masses are plotted as functions temperature for two values of sigma mass.

In Figure 2, the sigma and pion masses are plotted as functions of temperature at coupling constant = 3.55, $g = 4.84$, and $\mu_B = 270$ MeV. We note that the behavior is not changed with increasing coupling constant $g$. The critical temperature decreases with increasing coupling constant $g$.

In Figure 3, we examine the effect of sigma mass on the behavior of the effective sigma and pion masses and the critical temperature in the presence of baryonic chemical potential. We note that the behavior is unchanged with strong increase in the sigma mass while the critical temperature shifted to higher value from $T_c = 94$ MeV to $T_c = 140$ MeV corresponding to $m_\sigma = 500$ MeV and $m_\sigma = 700$ MeV, respectively. The similar effect is obtained in [17] in the absence of baryonic chemical potential. Also, we obtained the similar effect using the logarithmic potential where the same technique is used [28].

Next, we need to examine the behavior of the pressure at finite temperature. In Figure 4, the pressure is plotted as a function of temperature in the absence and presence of the chemical potential. We note that the curve increases continuously with increasing temperature. The values of the pressure at lower-values of temperature are not sensitive in comparison with the values of the pressure at the higher values of temperature. Hence the effect of largest values of temperature has more effect on the pressure. By increasing baryon chemical potential up to $\mu_b = 270$ MeV, we note the pressure shifted to higher values. This means the effect of the baryonic chemical potential has more effect on the pressure at higher values of temperature. Berger and Christov [29] found that the pressure increases with increasing temperature in hot medium using the NJL model in the mean-field approximation. Bowman and Kapusta [25] studied the pressure as a function of baryon chemical potential for different values of temperatures. They found that the pressure increases with increasing chemical potential. Hence, the present results of the pressure are in agreement with [25, 29].

Next, we need to examine the energy density in the presence of the baryon chemical potential. In Figure 5, the energy density is plotted as a function of temperature at $\mu_b = 0$ and $\mu_b = 270$ MeV; we note that the energy density increases smoothly with increasing temperature at $\mu_b = 0$ and $\mu_b =$

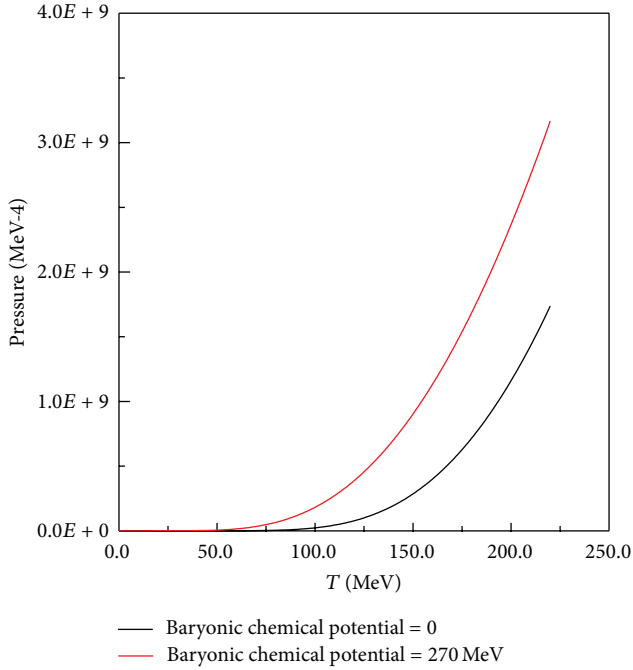

Figure 4: The pressure is plotted as a function of temperature for two values of $\mu_B = 0$ and $\mu_b = 270$ MeV.

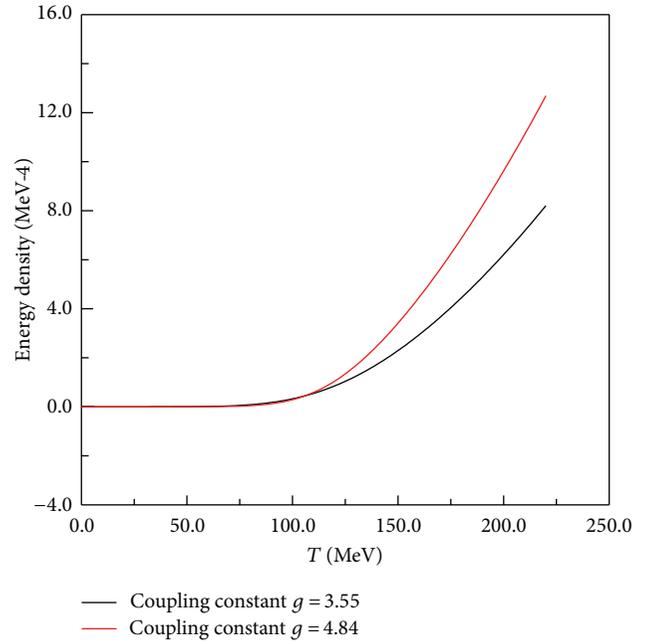

Figure 6: The energy density is plotted as function of temperature for two values of coupling constant $g$ at $\mu_b = 186$ MeV.

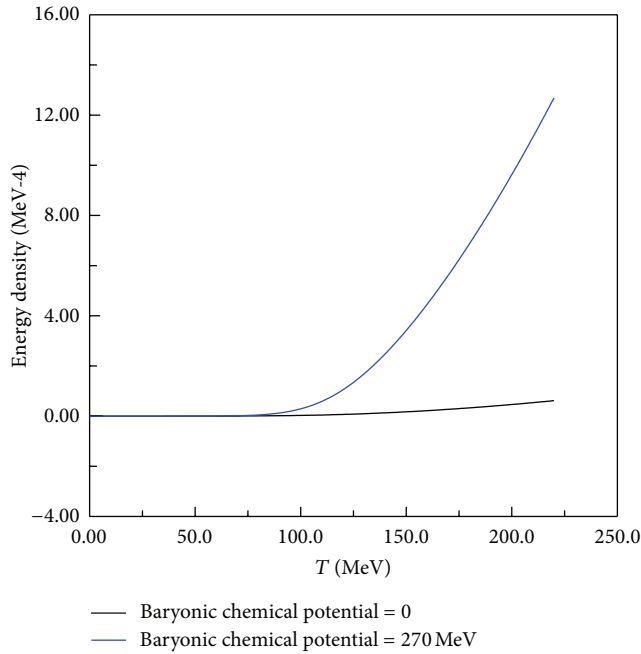

Figure 5: The energy density is plotted as a function of temperature for two values of $\mu_B = 0$ and $\mu_b = 270$ MeV.

270 MeV. The energy density is not sensitive to the baryonic chemical potential up to $T = 75$ MeV. Therefore, the effect of the baryon chemical potential appears at higher values of temperature. An increase in the baryonic chemical potential leads to increasing energy density as in **Figure 6**. Berger and Christov [29] studied the energy density in the NJL model and found that the energy density increases with increasing temperature up to $T = 150$ MeV at finite density. Bowman and Kapusta [25] studied the energy density in the linear sigma model. They found that energy density increases as a continuous curve with increasing temperature for a vacuum pion mass of 200 MeV. This indicated that the phase transition is crossover. Therefore, the present result refers to the same conclusion that the phase transition is crossover at pion mass equals 140 MeV. In the present quark model, we included the fermion sector term which is ignored in other works such as in [10, 12, 15]. So we need to examine the effect of this factor on the energy density. In **Figure 6**, the energy density is plotted as a function of temperature at the coupling constant $g = 3.55$ and $g = 4.84$, and $\mu_b = 186$ MeV. We note that the energy density of quarks decreases with increasing the coupling content due to the bound state is more tight.

## 5. Summary and Conclusion

In this work, we investigated the effect of the baryon chemical potential on the meson properties, the pressure, and the energy density at finite temperature. The effective mesonic potential is calculated by using the N-midpoint method at finite of chemical potential. We found that the behavior of the meson properties is in good agreement in comparison with other models. The pressure and energy density are examined as functions of temperature at finite temperature. A comparison is presented with other works. In addition, we included the fermion sector in the linear sigma model. This sector is ignored in many previous studies such as in [10, 12, 15].

We conclude that the present technique successfully predicts the behavior of the meson properties, the critical point temperature, and the energy density in comparison with other works at finite chemical potential and avoid the difficulty that previously mentioned in [17].

## Appendix

In this appendix, we write the basic steps of the $N$-midpoint rule to calculate the integral $M = \int_a^b f(x)dx$. We divide the interval $[a, b]$ into $N$ subintervals of length

$$\Delta x = \frac{b-a}{N}, \tag{A.1}$$

and the midpoint of the $i$th interval $[x_i, x_{i-1}]$ is

$$c_i = a + \left(i - \frac{1}{2}\right)\Delta x. \tag{A.2}$$

The $i$ the midpoint rectangle is the rectangle of height $f(c_i)$ over the subinterval $[x_i, x_{i-1}]$. This rectangle signed area

$$f(c_i)\Delta x. \tag{A.3}$$

Thus, $M$ is equal to the sum of the signed areas of these rectangles

$$M = \sum_{i=1}^{n} f(c_i)\Delta x, \tag{A.4}$$

by increasing the parameter $n$, we get a good accuracy for the numerical integration.

## Conflict of Interests

The author declares that there is no conflict of interests regarding the publication of this paper.

## References

[1] M. Fukugita, "Lattice quantum chromodynamics with dynamical quarks," *Nuclear Physics B*, vol. 4, pp. 105–129, 1988.

[2] M. Gell-Mann and M. Levy, "The axial vector current in beta decay," *Il Nuovo Cimento*, vol. 16, no. 4, pp. 705–726, 1960.

[3] M. C. Birse and M. K. Banerjee, "Chiral model for nucleon and delta," *Physical Review D*, vol. 31, no. 1, pp. 118–127, 1985.

[4] W. Broniowski and B. Golli, "Approximating chiral quark models with linear sigma-models," *Nuclear Physics A*, vol. 714, no. 3-4, pp. 575–588, 2003.

[5] M. Abu-Shady, "Effect of the A-term on the nucleon properties in the extended linear sigma model," *International Journal of Theoretical Physics*, vol. 48, no. 4, pp. 1110–1121, 2009.

[6] M. Abu-Shady, "Effect of coherent-pair approximation on nucleon properties in the extended linear sigma model," *Acta Physica Polonica B*, vol. 40, no. 8, pp. 2225–2238, 2009.

[7] M. Abu-Shady, "Effect of logarithmic mesonic potential on nucleon properties," *Modern Physics Letters A*, vol. 24, no. 20, pp. 1617–1629, 2009.

[8] M. Abu-Shady and M. Rashdan, "Effect of a logarithmic mesonic potential on nucleon properties in the coherent-pair approximation," *Physical Review C*, vol. 81, no. 1, Article ID 015203, 8 pages, 2010.

[9] M. Abu-Shady, "Chiral logarithmic quark model of $N$ and $\triangle$ with an A-term in the meanfield approximation," *International Journal of Modern Physics A*, vol. 26, no. 2, p. 235, 2011.

[10] N. Petropoulos, "Linear sigma model and chiral symmetry at finite temperature," *Journal of Physics G*, vol. 25, no. 11, pp. 2225–2241, 1999.

[11] G. Amelino-Camelia and S.-Y. Pi, "Self-consistent improvement of the finite-temperature effective potential," *Physical Review D*, vol. 47, no. 6, pp. 2356–2362, 1993.

[12] Y. Nemoto, K. Naito, and M. Oka, "Effective potential of the O($N$) linear sigma-model at finite temperature," *European Physical Journal A*, vol. 9, no. 2, pp. 245–259, 2000.

[13] M. Abu-Shady, "Meson properties at finite temperature in the linear sigma model," *International Journal of Theoretical Physics*, vol. 49, no. 10, pp. 2425–2436, 2010.

[14] M. Abu-Shady and H. Mansour, "Quantized linear sigma model at finite temperature and nucleon properties," *Physical Review C*, vol. 85, Article ID 055204, 2012.

[15] J. T. Lenaghan and D. H. Rischke, "The O($N$) model at finite temperature : renormalization of the gap equations in Hartree and large-$N$ approximations," *Journal of Physics G*, vol. 26, no. 4, p. 431, 2000.

[16] H. C. G. Caldas, A. L. Mota, and M. C. Nemes, "The chiral fermion meson model at finite temperature," *Physical Review D*, vol. 63, no. 5, Article ID 056011, 2001.

[17] M. Abu-Shady, "N-midpoint rule for calculating the effective mesonic potential at finite temperature," *International Journal of Theoretical Physics*, vol. 52, pp. 1165–1174, 2013.

[18] O. Scavenius, A. Mocsy, I. N. Mishustin, and D. H. Rischke, "Chiral phase transition within effective models with constituent quarks," *Physical Review C*, vol. 64, no. 4, Article ID 045202, 2001.

[19] T. Brauner, "Spontaneous symmetry breaking in the linear sigma model at finite chemical potential," *Physical Review D*, vol. 74, no. 8, Article ID 085010, 13 pages, 2006.

[20] N. Bilic and H. Nikolic, "Chiral-symmetry restoration in the linear sigma model at nonzero temperature and baryon density," *European Physical Journal C*, vol. 6, no. 3, pp. 513–523, 1999.

[21] H. Mao, N. Petropoulos, S. Shu, and W.-Q. Zhao, "The linear sigma model at a finite isospin chemical potential," *Journal of Physics G*, vol. 32, no. 11, pp. 2187–2198, 2006.

[22] D. Jogerman, "Investigation of a modified mid-point quadrature formula," *Mathematics of Computation*, vol. 20, pp. 79–89, 1966.

[23] F. Stetter, "On a generalization of the midpoint rule," *Mathematics of Computation*, vol. 22, pp. 661–663, 1968.

[24] M. Nahrgang and M. Bleiher, "Fluid dynamics with a critical point," *Acta Physica Polonica B*, vol. 2, p. 405, 2009.

[25] E. S. Bowman and J. I. Kapusta, "Critical points in the linear sigma model with quarks," *Physical Review C*, vol. 79, no. 1, Article ID 015202, 7 pages, 2009.

[26] J.-W. Chen, H. Kohyama, and U. Raha, "Model analysis of thermal UV-cutoff effects on the chiral critical surface at finite temperature and chemical potential," *Physical Review D*, vol. 83, no. 9, Article ID 094014, 2011.

[27] H. Hansen, W. M. Alberico, A. Beraudo, A. Molinari, M. Nardi, and C. Ratti, "Mesonic correlation functions at finite

temperature and density in the Nambu-Jona-Lasinio model with a Polyakov loop," *Physical Review D*, vol. 75, no. 6, Article ID 065004, 2007.

[28] M. Abu-shady, "A new technique for the calculation of effective mesonic potential at finite temperature in the logarithmic quark-sigma model," *Journal of Fractional Calculus and Applications*, vol. 3(s), p. 6, 2012.

[29] J. Berger and C. V. Christov, "Phase transition and the nucleon as a soliton in the Nambu-Jona-Lasinio model at finite temperature and density," *Nuclear Physics A*, vol. 609, no. 4, pp. 537–561, 1996.

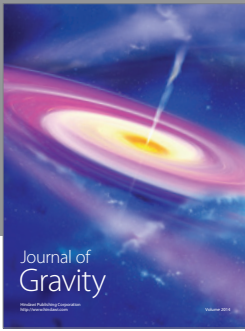
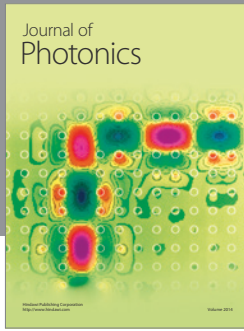
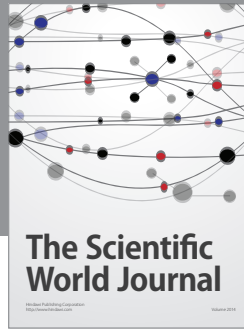
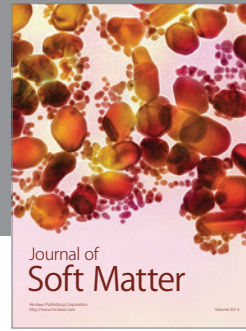
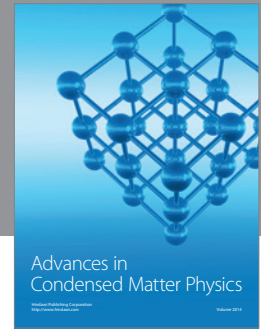
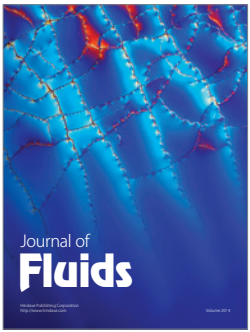
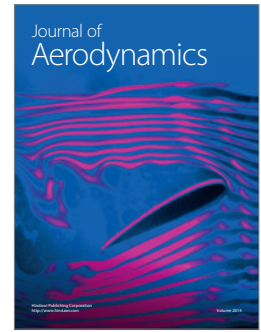
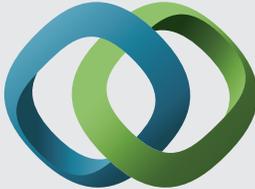

Submit your manuscripts at
http://www.hindawi.com

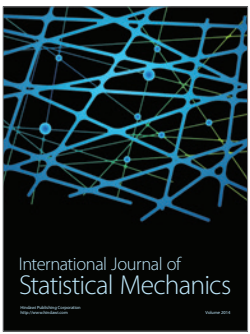
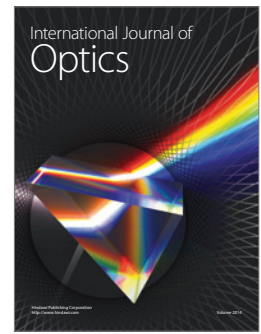
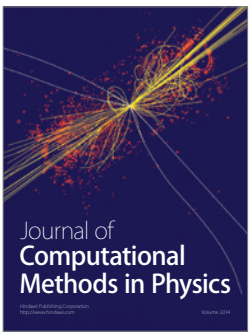
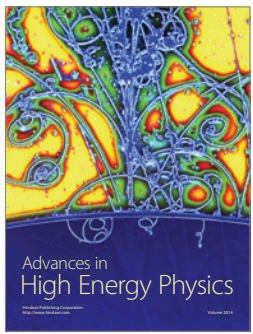
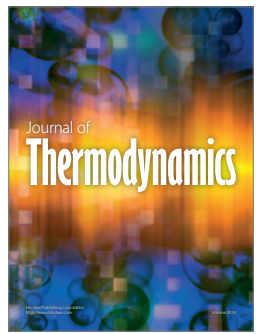
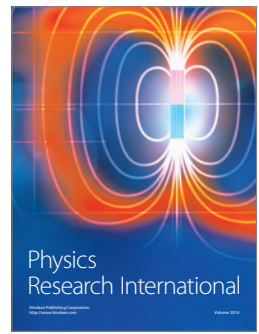
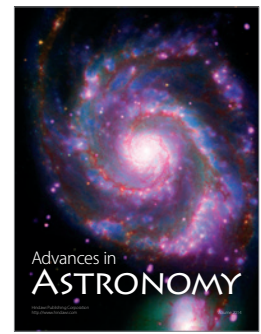
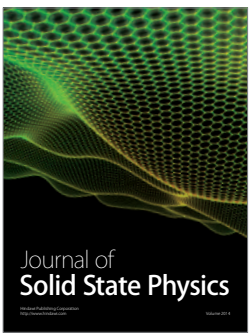
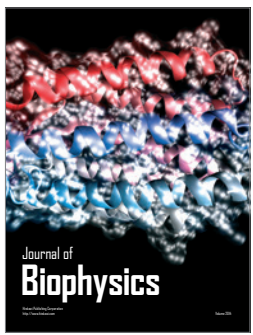
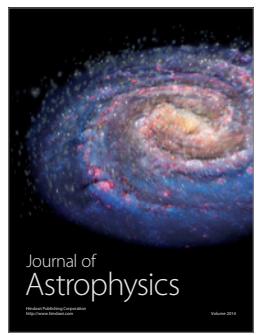
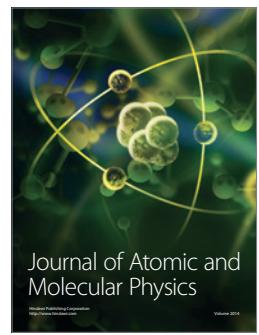
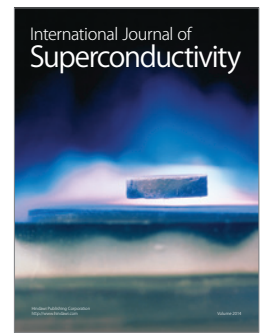